\documentclass[runningheads]{llncs}
\usepackage{amssymb}
\usepackage{amsmath}
\usepackage{graphicx}
\usepackage{color}
\usepackage[algo2e,ruled,vlined]{algorithm2e}

\newtheorem{observation}[theorem]{Observation}

\usepackage{color}
\definecolor{red}{rgb}{1,0,0}

\begin{document}

\pagestyle{headings}
\mainmatter

\title{Graphs with Extremal Connected \\Forcing Numbers}
\titlerunning{Extremal Connected Forcing Numbers}

\author{Boris Brimkov, Caleb C. Fast, and Illya V. Hicks}
\authorrunning{B. Brimkov, C. C. Fast, I. V. Hicks}
\institute{
Department of Computational \& Applied Mathematics, Rice University,\\ Houston, TX 77005, USA\\
\email{boris.brimkov@rice.edu, caleb.c.fast@rice.edu, ivhicks@rice.edu}
}

\maketitle

\begin{abstract}
Zero forcing is an iterative graph coloring process where at each discrete time step, a colored vertex with a single uncolored neighbor forces that neighbor to become colored. The zero forcing number of a graph is the cardinality of the smallest set of initially colored vertices which forces the entire graph to eventually become colored. Connected forcing is a variant of zero forcing in which the initially colored set of vertices induces a connected subgraph; the analogous parameter of interest is the connected forcing number. In this paper, we characterize the graphs with connected forcing numbers 2 and $n-2$. Our results extend existing characterizations of graphs with zero forcing numbers 2 and $n-2$; we use combinatorial and graph theoretic techniques, in contrast to the linear algebraic approach used to obtain the latter. We also present several other structural results about the connected forcing sets of a graph.
\smallskip


\end{abstract}

\section{Introduction}
Zero forcing is an iterative graph coloring process where at each discrete time step, a colored vertex with a single uncolored neighbor forces that neighbor to become colored. The zero forcing number of a graph is the cardinality of the smallest set of initially colored vertices which forces the entire graph to be colored. Zero forcing was initially used to bound the maximum nullity of the family of symmetric matrices described by a graph \cite{AIM-Workshop}, but has since found a variety of uses in physics, logic circuits, coding theory, power network monitoring, and in modeling the spread of diseases and information in social networks; see \cite{zf_tw,quantum1,logic1,fallat,powerdom3,proptime1,zf_np,powerdom2} for more details. 

Connected forcing is a variant of zero forcing in which the initially colored set of vertices induces a connected subgraph. The connected forcing number of a graph is the cardinality of the smallest connected set of initially colored vertices which forces the entire graph to be colored. Various structural and computational aspects of connected forcing have been investigated in \cite{brimkov,brimkov2,CF_paper}. The connected forcing number bounds parameters such as the maximum nullity, path cover number, and power domination number (cf. \cite{brimkov,CF_paper}); it can also potentially be applied to power network monitoring and modeling propagation of information (cf. \cite{brimkov2}). Other variants of zero forcing, such as positive semidefinite zero forcing \cite{Barioli,positive_zf2,proptime2}, fractional zero forcing \cite{fractional_zf}, and signed zero forcing \cite{signed_zf} have also been studied; see also \cite{butler,kenter} and the bibliographies therein. These variants are typically obtained by modifying the zero forcing color change rule, or adding certain restrictions to the structure of a forcing set. 

Computing the zero forcing number and connected forcing number of a graph are both NP-complete problems \cite{aazami,brimkov2}; thus, approaches addressing the complexity of these problems include developing closed formulas, characterizations, and bounds for the forcing numbers of graphs with special structure; such results are obtained in \cite{AIM-Workshop,benson,brimkov,brimkov2,Eroh,Huang,Meyer}. In particular, graphs whose zero forcing number equals $1$, $2$, and $n-1$  have been characterized in \cite{row}, and graphs whose zero forcing number equals $n-2$ have been characterized in \cite{AIM-Workshop}; similarly, graphs whose connected forcing number equals 1 and $n-1$ have been characterized in \cite{brimkov}. In this paper, we extend these results by characterizing graphs whose connected forcing numbers are 2 and $n-2$. Other related characterizations have been derived for graphs whose minimal rank is two \cite{barrett,barrett1} and three \cite{barrett2}, graphs whose positive semi-definite matrices have nullity at most two \cite{holst1}, three-connected graphs whose maximum nullity is at most three \cite{holst2}, and graphs for which the maximum multiplicity of an eigenvalue is two \cite{johnson}. Many of these characterizations have been obtained using linear algebraic approaches; in contrast, we employ novel combinatorial and graph theoretic techniques which make use of the vertex connectivity of a graph and the connectedness of its forcing set. We also present several other structural results, and introduce a generalization of zero forcing whose further study could be of independent interest.

The paper is organized as follows. In the next section, we recall some graph theoretic notions, specifically those related to zero forcing. In Section 3, we characterize graphs with connected forcing numbers 2 and $n-2$, and present several other structural results about connected forcing sets. We conclude with some final remarks and open questions in Section 4.

\section{Preliminaries}

\subsection{Graph theoretic notions}
A graph $G=(V,E)$ consists of a vertex set $V$ and an edge set $E$ of two-element subsets of $V$. The \emph{order} and \emph{size} of $G$ are denoted by $n=|V|$ and $m=|E|$, respectively. Two vertices $v,w\in V$ are \emph{adjacent}, or \emph{neighbors}, if $\{v,w\}\in E$. If $v$ is adjacent to $w$, we write $v\sim w$; otherwise, we write $v\not\sim w$. The \emph{degree} of a vertex $v$ in $G$, denoted $d(v;G)$, is the number of neighbors $v$ has in $G$; the dependence on $G$ can be ommitted when it is clear from the context. The minimum degree and maximum degree of $G$ are denoted by $\delta(G)$ and $\Delta(G)$, respectively. A \emph{leaf}, or \emph{pendant}, is a vertex with degree 1. An \emph{isolated vertex} or \emph{isolate} is a vertex with degree 0; such a vertex will also be called a \emph{trivial (connected) component} of $G$. Given $S \subset V$, the \emph{induced subgraph} $G[S]$ is the subgraph of $G$ whose vertex set is $S$ and whose edge set consists of all edges of $G$ which have both endpoints in $S$. An isomorphism between graphs $G_1$ and $G_2$ will be denoted by $G_1\simeq G_2$. The number of connected components of a graph will be denoted by $\emph{comp}(G)$. 

A \emph{separating set} of $G$ is a set of vertices which, when removed, increases the number of connected components in $G$. A \emph{cut vertex} is a separating set of size one. The \emph{vertex connectivity} of $G$, denoted $\kappa(G)$, is the largest number such that $G$ remains connected whenever fewer than $\kappa(G)$ vertices of $G$ are removed; a disconnected graph has vertex connectivity zero. A \emph{cut edge} is an edge which, when removed, increases the number of connected components of $G$. A \emph{biconnected component}, or \emph{block}, of $G$ is a maximal subgraph of $G$ which has no cut vertices; $G$ is \emph{biconnected} if it has no cut vertices. An \emph{outer block} is a block which contains at most one cut vertex of $G$. A \emph{trivial block} is a block with two vertices, i.e., a cut edge of $G$.

The \emph{disjoint union} of sets $S_1$ and $S_2$, denoted $S_1\dot\cup S_2$, is a union operation that indexes the elements of the union set according to which set they originated in; the \emph{disjoint union} of two graphs $G_1=(V_1,E_1)$ and $G_2=(V_2,E_2)$, denoted $G_1\dot\cup G_2$, is the graph $(V_1\dot\cup V_2,E_1\dot\cup E_2)$. The \emph{join} of two graphs $G_1$ and $G_2$, denoted $G_1\lor G_2$, is the graph obtained from $G_1\dot\cup G_2$ by adding an edge from each vertex of $G_1$ to each vertex of $G_2$. The \emph{complement} of a graph $G=(V,E)$ is the graph $G^c=(V,E^c)$. A \emph{complete} graph is denoted $K_n$, and a \emph{complete bipartite} graph, denoted $K_{p,q}$ is the complement of $K_p\dot\cup K_q$ (we may allow these indices to equal 0, in which case $K_{n,0}\simeq K_{0,n}\simeq \dot\bigcup_{i=1}^n K_1$). A graph with no edges will be called an \emph{empty graph}; a path on $n$ vertices will be denoted $P_n$. If $\mathcal{F}$ is a set of graphs, a graph is $\mathcal{F}$\emph{-free} if it does not contain $F$ as an induced subgraph for each $F\in\mathcal{F}$. For other graph theoretic terminology and definitions, we refer the reader to \cite{bondy}.

\subsection{Zero forcing}
Given a graph $G=(V,E)$ and a set $S \subset V$ of initially colored vertices, the \emph{color change rule} dictates that at each integer-valued time step, a colored vertex $u$ with a single uncolored neighbor $v$ \emph{forces} that neighbor to become colored; such a \emph{force} is denoted $u\rightarrow v$. 
The \emph{derived set} of $S$ is the set of colored vertices obtained after the color change rule is applied until no new vertex can be forced; it can be shown that the derived set of $S$ is uniquely determined by $S$ (see \cite{AIM-Workshop}). A \emph{zero forcing set} is a set whose derived set is all of $V$; the \emph{zero forcing number} of $G$, denoted $Z(G)$, is the minimum cardinality of a zero forcing set. 

A \emph{chronological list of forces} of $S$ is a sequence of forces applied to obtain the derived set of $S$ in the order they are applied; there can also be initially colored vertices which do not force any vertex. 
Generally, the chronological list of forces is not uniquely determined by $S$; for example, it may be possible for several colored vertices to force an uncolored vertex at a given step. 
A \emph{forcing chain} for a chronological list of forces is a maximal sequence of vertices $(v_1,\ldots,v_k)$ such that $v_i\rightarrow v_{i+1}$ for $1\leq i\leq k-1$. It may be possible for an initially colored vertex not to force any vertex. If a vertex forces another vertex at some step of the forcing process, then it cannot force a second vertex at a later step, since that would imply it had two uncolored neighbors when it forced for the first time. Thus, each forcing chain induces a distinct path in $G$, one of whose endpoints is an initially colored vertex and the rest of whose vertices are uncolored at the initial time step; we will say the initially colored vertex \emph{initiates} the forcing chain. The set of all forcing chains for a chronological list of forces is uniquely determined by the chronological list of forces and forms a path cover of $G$.

A \emph{connected zero forcing set} of $G$ is a zero forcing set of $G$ which induces a connected subgraph. The \emph{connected zero forcing number} of $G$, denoted $Z_c(G)$, is the cardinality of a minimum connected zero forcing set of $G$. For short, we may refer to these as \emph{connected forcing set} and \emph{connected forcing number}. Note that a disconnected graph cannot have a connected forcing set.

\section{Graphs with extremal connected forcing numbers}


Polynomial time algorithms and closed-form expressions have been derived for computing the connected forcing numbers of special classes of graphs, including trees, unicyclic graphs, grid graphs, sun graphs, and several other families (cf. \cite{brimkov,brimkov2}). Conversely, a complete characterization of graphs having a particular connected forcing number can be obtained through a combinatorial case analysis. For example, it is easy to see that $Z_c(G)=1$ if and only if $G$ is a path $P_n$. Moreover, Brimkov and Davila \cite{brimkov} gave the following characterization of graphs with connected forcing number $n-1$.
\begin{theorem}\emph{\cite{brimkov}}
\label{cfn1}
$Z_c(G) = n-1$ if and only if $G\simeq K_n$, $n\geq 2$, or $G\simeq K_{1,n-1}$, $n\geq 4$.
\end{theorem}
In what follows, we extend these results by characterizing graphs for which $Z_c(G)=2$ and $Z_c(G)=n-2$.

\subsection{Graphs with $Z_c(G)=2$}
In this section, we will characterize all graphs with connected forcing number 2. We first recall some definitions and previous results.

\begin{definition}
\label{def_pendant_tree}
A \emph{pendant path attached to vertex} $v$ in graph $G=(V,E)$ is a set $P\subset V$ such that $G[P]$ is a connected component of $G-v$ which is a path, one of whose ends is adjacent to $v$ in $G$. The neighbor of $v$ in $P$ will be called the \emph{base} of the path, and $p(v)$ will denote the number of pendant paths attached to $v\in V$. 
\end{definition}

\begin{definition}
Let $G=(V,E)$ be a connected graph. Define 
\begin{eqnarray*}
R_1(G)&=&\{v\in V: \emph{comp}(G-v)=2, \; p(v)=1\}\\
R_2(G)&=&\{v\in V: \emph{comp}(G-v)=2, \; p(v)=0\}\\
R_3(G)&=&\{v\in V: \emph{comp}(G-v)\geq 3\}\\
\mathcal{L}(G)&=&\bigcup_{v\in V}\{\text{all-but-one bases of pendant paths attached to } v\}\\
M(G)&=&R_2(G)\cup R_3(G)\cup \mathcal{L}(G). 
\end{eqnarray*} 
When there is no scope for confusion, the dependence on $G$ will be omitted.
\end{definition}

\begin{lemma} \emph{\cite{brimkov2}}
\label{MR_lemma}
Let $G=(V,E)$ be a connected graph different from a path and $R$ be an arbitrary connected forcing set of $G$. Then $M(G)\subset R$.
\end{lemma}

\begin{definition}
\label{parallel_path_def}
A graph $G=(V,E)$ is a \emph{graph of two parallel paths specified by $V_1$ and $V_2$} if $G\not\simeq P_n$, and if $V$ can be partitioned into nonempty sets $V_1$ and $V_2$ such that $P_1:=G[V_1]$ and $P_2:=G[V_2]$ are paths, and such that $G$ can be drawn in the plane in such a way that $P_1$ and $P_2$ are parallel line segments, and the edges between $P_1$ and $P_2$ (drawn as straight line segments) do not cross; such a drawing of $G$ is called a \emph{standard drawing}. In a standard drawing of $G$, fix an ordering of the vertices of $P_1$ and $P_2$ that is increasing in the same direction for both paths. In this ordering, let $\emph{first}(P_i)$ and $\emph{last}(P_i)$ respectively denote the first and last vertices of $P_i$ for $i=1,2$. The sets $\{\emph{first}(P_1),\emph{first}(P_2)\}$ and $\{\emph{last}(P_1),\emph{last}(P_2)\}$ will be referred to as \emph{ends} of $G$.

\end{definition}

Note that if $G$ is a graph of two parallel paths, there may be several different partitions of $V$ into $V_1$ and $V_2$ which satisfy the conditions above. For example, let $G=(\{1,2,3,4,5\},\{\{1,2\},\{2,3\},\{3,4\},\{4,5\},\{5,1\}\})$ be a cycle on 5 vertices. Then $G$ is a graph of two parallel paths that can be specified by $V_1=\{1\}$ and $V_2=\{2,3,4,5\}$, as well as by $V_1=\{1,2,3\}$ and $V_2=\{4,5\}$.

Graphs of two parallel paths were introduced by Johnson et al. \cite{johnson} in relation to graphs with maximum nullity 2. They were also used by Row \cite{row} in the following characterization.

\begin{theorem}\emph{\cite{row}}
\label{zf2}
$Z(G)=2$ if and only if $G$ is a graph of two parallel paths.
\end{theorem}
The following observation regarding the result of Theorem \ref{zf2} is readily verifiable (and has been noted in \cite{row}).
\begin{observation}
\label{aux_zf2_obs}
Either end of a graph on two parallel paths is a zero forcing set. Conversely, if $Z(G)=2$, the two forcing chains associated with a minimum zero forcing set induce a specification of $G$ as a graph on two parallel paths.
\end{observation}
The following observation follows from the definition of forcing vertices. 
\begin{observation}
\label{obs_delta}
Every minimum zero forcing set and every minimum connected forcing set contains a vertex together with all-but-one of its neighbors.
\end{observation}
Finally, let $L(G)$ denote the set of leaves of $G$; we recall a result of Brimkov and Davila \cite{brimkov} relating $Z_c(G)$ to $|L(G)|$.
\begin{lemma}\emph{\cite{brimkov}}
\label{leaf_lemma}
For any connected graph $G$ different from a path, $Z_c(G)\geq |L(G)|$.
\end{lemma}
We now prove the main result of this section.
\begin{theorem}
\label{thm_cf2}
$Z_c(G)=2$ if and only if $G$ belongs to the family of graphs described in Figures \ref{fig_cf2case1} and \ref{fig_cf2case2}.
\end{theorem}
\proof
Let $G=(V,E)$ be a graph with $Z_c(G)=2$. Since $Z(G)=1$ if and only if $Z_c(G)=1$, and since $2=Z_c(G)\geq Z(G)$, it follows that $Z(G)=2$. Thus, by Theorem~\ref{zf2}, $G$ is a graph of two parallel paths.
Fix some partition of $V$ into $V_1$ and $V_2$ which satisfies Definition \ref{parallel_path_def}, fix a standard drawing of $G$ based on that partition, and fix a vertex ordering as specified in Definition \ref{parallel_path_def}. From Lemma~\ref{leaf_lemma}, it follows that $G$ has 0, 1, or 2 leaves. 


\begin{claim}
\label{claim_two_edges}
Let $G$ be a graph of two parallel paths with $Z_c(G)=2$. Then, there are at least two edges between the two parallel paths of $G$. 
\end{claim}
\proof
If there are no edges between the two parallel paths of $G$, then $G$ is disconnected, and cannot have a connected forcing set. If there is one edge between the two parallel paths, then $G$ is either isomorphic to a path (and is hence not a graph of two parallel paths), or has more than two leaves (and hence $Z_c(G)>2$ by Lemma~\ref{leaf_lemma}). Thus, there must be at least two edges with one endpoint in $V_1$ and the other in $V_2$.
\qed
\vspace{9pt}
\noindent We will now consider several cases based on the number and position of the leaves in $G$.

\begin{claim}
\label{top_of_proof}
Let $G$ be a graph of two parallel paths which has 0 leaves, 1 leaf, or 2 leaves which belong to the same end of $G$. Then, $Z_c(G)=2$, and $G$ belongs to the family of graphs described in Figure \ref{fig_cf2case1}.
\begin{figure}[ht!]
\begin{center}
\includegraphics[scale=0.7]{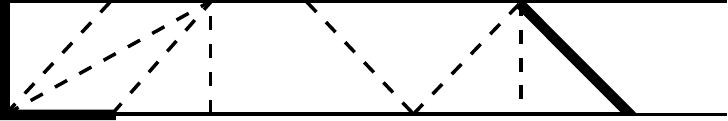}
\caption{A graph of two parallel paths with 0 leaves, 1 leaf, or 2 leaves which belong to the same end of the graph; solid lines represent paths of arbitrary (possibly zero) length; bold lines represent mandatory single edges; dashed lines represent any configuration of non-intersecting edges between the parallel paths.}
\label{fig_cf2case1}
\end{center}
\end{figure}
\end{claim}

\proof
Without loss of generality, suppose $L(G)\subseteq \{\emph{last}(V_1),\emph{last}(V_2)\}$. Let $V_1'\subseteq V_1$ and $V_2'\subseteq V_2$ be maximal sets of vertices which do not belong to pendant paths of $G$. By Claim \ref{claim_two_edges}, there are at least two distinct edges with one endpoint in $V_1'$ and the other in $V_2'$; thus, it follows that at least one of the paths $G[V_1']$ and $G[V_2']$ must have length greater than zero. Moreover, by Observation \ref{aux_zf2_obs}, and since $\emph{first}(V_1)$ and $\emph{first}(V_2)$ are adjacent, it follows that $\{\emph{first}(V_1),\emph{first}(V_2)\}$ is a connected forcing set; thus, $Z_c(G)=2$. This is the family of graphs illustrated in Figure~\ref{fig_cf2case1}.
\qed


\begin{claim}
\label{claim_samepath}
Let $G$ be a graph of two parallel paths that has 2 leaves which belong to the same path and different ends of $G$. Then $Z_c(G)=2$ if and only if $G$ belongs to the family of graphs described in Figure \ref{fig_cf2case2}.
\begin{figure}[ht!]
\begin{center}
\includegraphics[scale=0.7]{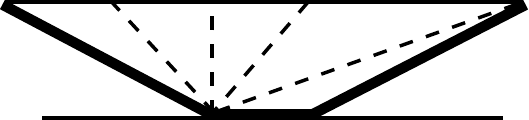}
\caption{A graph of two parallel paths with 2 leaves which belong to different ends of the graph; solid lines represent paths of arbitrary (possibly zero) length; bold lines represent mandatory single edges; dashed lines represent a configuration of non-intersecting edges between the parallel paths, all of which are incident to the same vertex in the path containing the mandatory single edge.}
\label{fig_cf2case2}
\end{center}
\end{figure}
\end{claim}

\proof
Without loss of generality, suppose $L(G)=\{\emph{first}(V_1),\emph{last}(V_1)\}$. Let $H_1$ be the pendant path containing $\emph{first}(V_1)$, and $u_1$ be the vertex to which $H_1$ is attached; let $H_2$ be the pendant path containing $\emph{last}(V_1)$, and $u_2$ be the vertex to which $H_2$ is attached. Let $V_1'=V_1\backslash(H_1\cup H_2)$. 

Suppose first that $|V_1'|=1$. Then $u_1=u_2$, and $H_1$ and $H_2$ are both attached to $u_1$; thus, by Lemma \ref{MR_lemma}, $u_1$ and some neighbor $z$ of $u_1$ in $H_1$ or $H_2$ must be contained in every minimum connected forcing set of $G$. However, a set containing only $u_1$ and $z$ is not forcing, since $u_1$ has at least two uncolored neighbors outside $H_1\cup H_2$; this is a contradiction. 

Suppose next that $|V_1'|\geq 3$, and let $R=\{r_1,r_2\}$ be a connected forcing set of $G$. By Observation \ref{obs_delta}, and since neither leaf of $G$ together with its neighbor forms a forcing set, it follows that at least one of $r_1$ and $r_2$ has degree 2. Without loss of generality, let $r_1$ be a vertex of degree 2. If $r_1$ is contained in $H_i$, for $i\in\{1,2\}$, then no vertex outside $H_i\cup\{u_i\}$ can be forced by $R$. Similarly, if $r_1$ is contained in $V_1'$, then no vertex outside $V_1'$ can be forced by $R$, and if $r_1$ is contained in $V_2'$, then no vertex outside $V_2'\cup\{u_1,u_2\}$ can be forced by $R$. Thus, the assumption that $|V_1'|\geq 3$ leads to a contradiction, so it follows that $|V_1'|=2$, i.e., $V_1'=\{u_1,u_2\}$. Recall that by Claim \ref{claim_two_edges}, each of $u_1$ and $u_2$ must be adjacent to at least one vertex of $V_2$ -- namely, $\emph{first}(V_2)$ and $\emph{last}(V_2)$, respectively.


Suppose first that both $u_1$ and $u_2$ are adjacent to two or more vertices of $V_2$. Let $v_1$ and $v_2$ respectively be the neighbors of $u_1$ and $u_2$ in $V_2$ which are respectively closest to $\emph{first}(V_2)$ and $\emph{last}(V_2)$ in $G[V_2]$; $v_1$ and $v_2$ could possibly be the same vertex. Let $S_1$, $S_2$, and $S_3$ respectively be the sets of vertices between $\emph{first}(V_2)$ and $v_1$, $v_1$ and $v_2$, and $v_2$ and $\emph{last}(V_2)$, inclusively (where ``between'' refers to the vertex ordering of $G$, i.e., to the position of the vertices in the path $G[V_2]$). As shown in the case where $|V_1'|\geq 3$, the degree 2 vertex $r_1$ cannot be contained in $H_i$, $i\in\{1,2\}$. Similarly, if $r_1$ is contained in $S_1$, $S_2$, or $S_3$, then, respectively, no vertex outside $S_1\cup\{u_1\}$, $S_2$, and $S_3\cup\{u_2\}$ can be forced by $R$. Once again, it follows that no set consisting of a degree 2 vertex and one of its neighbors can force all of $G$, a contradiction.


Thus, one of $u_1$ and $u_2$, say $u_1$, must be adjacent to a single vertex of $V_2$, namely $\emph{first}(V_2)$. Then $\{u_2,\emph{last}(V_2)\}$ is a connected forcing set, since $\emph{last}(V_2)$ can initiate a forcing chain passing through all vertices in $V_2$ and eventually forcing $u_1$; then $u_1$ and $u_2$ will be able to force $H_1$ and $H_2$, respectively. This is the family of graphs illustrated in Figure \ref{fig_cf2case2}.
\qed

\begin{claim}
Let $G$ be a graph of two parallel paths which has 2 leaves which belong to different paths and different ends of $G$. Then $Z_c(G)=2$ if and only if $G$ (can be respecified as a graph which) belongs to the family of graphs described in Figure \ref{fig_cf2case2}.
\end{claim}
\proof
Without loss of generality, suppose $L(G)=\{\emph{first}(V_1),\emph{last}(V_2)\}$. Let $H_1$ be the pendant path containing $\emph{first}(V_1)$, and $u_1$ be the vertex to which $H_1$ is attached; let $H_2$ be the pendant path containing $\emph{last}(V_2)$, and $u_2$ be the vertex to which $H_2$ is attached. Let $V_1'=V_1\backslash H_1$ and $V_2'=V_2\backslash H_2$. Since $G$ is different from a single path, it cannot be the case that $|V_1'|=1$ and $|V_2'|=1$. 

Suppose $|V_1'|\geq 2$ and $|V_2'|\geq 2$, and let $R=\{r_1,r_2\}$ be a connected forcing set of $G$. By the same argument as in Claim \ref{claim_samepath}, one of $r_1$ and $r_2$, say $r_1$, must have degree 2; moreover, $r_1$ cannot be contained in $H_1$ or $H_2$. If $r_1$ is contained in $V_1'$, then no vertex outside $V_1'\cup\{u_2\}$ can be forced by $R$, a contradiction. By symmetry, $r_1$ also cannot be in $V_2'$. 

Thus, exactly one of $V_1'$ and $V_2'$ consists of a single vertex; without loss of generality, suppose $|V_1'|=1$ and $|V_2'|\geq 2$. Note then, that $u_1=\emph{last}(V_1)$, and all edges between $V_1'$ and $V_2'$ are incident to $u_1$. Let $w$ be the neighbor of $u_2$ in $V_2'$, which exists by the assumption that $|V_2'|\geq 2$. Then, the vertex partition $\widehat{V_1}=H_1\cup \{u_1,u_2\}\cup H_2$, $\widehat{V_2}=V\backslash \widehat{V_1}$ gives an alternate specification of $G$ as a graph of two parallel paths. In this specification, the two leaves of $G$ belong to the same path and different ends of $G$. Thus, by Claim \ref{claim_samepath}, $Z_c(G)=2$ if and only if $G$ belongs to the family of graphs described in Figure \ref{fig_cf2case2}. 
\qed
\vspace{9pt}
\noindent Since there are no other possible positions for the leaves of $G$, this concludes the proof of Theorem \ref{thm_cf2}.
\qed

\subsection{Graphs with $Z_c(G)=n-2$}

In this section, we will characterize all graphs with connected forcing number $n-2$. We begin by recalling a result regarding graphs with zero forcing number $n-2$.

\begin{theorem}\emph{\cite{AIM-Workshop}}
\label{thm_zfn2}
$Z(G) \geq n-2$ if and only if $G$ does not contain an induced subgraph isomorphic to any of the graphs in Figure \ref{fig_zfn2}.
\begin{figure}[ht!]
\begin{center}
\includegraphics[scale=0.35]{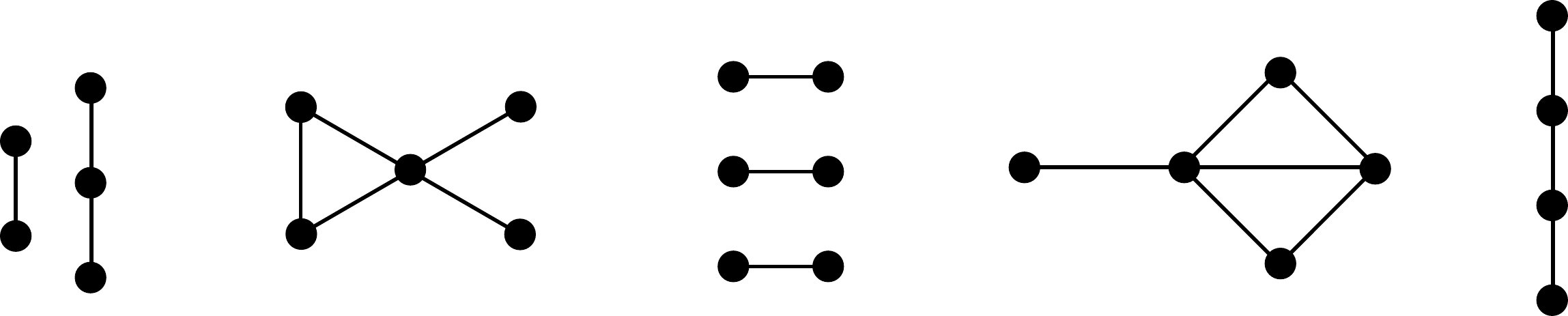}
\caption{Forbidden induced subgraphs for $Z(G)\geq n-2$; from left to right: $P_2\dot\cup P_3$, ``fish", $P_2\dot\cup P_2\dot\cup P_2$, ``dart", $P_4$.}
\label{fig_zfn2}
\end{center}
\end{figure}
\end{theorem}

Theorem  \ref{thm_zfn2} is a consequence of the following characterization of the graphs with minimum Hermitian rank at most 2, due to Barrett, van der Holst, and Loewy~\cite{barrett}. 

\begin{theorem}\emph{\cite{barrett}}
\label{thm_hmr2}
Given a graph $G=(V,E)$ with vertex set $V=\{1,\ldots,n\}$, let $H(G)$ be the set of all Hermitian $n\times n$ matrices $A=[a_{ij}]$ such that $a_{ij}\neq 0$ for $i\neq j$, if and only if $\{i,j\}\in E$ (and no restriction on $a_{ii}$). Let $\emph{hmr}(G)=\min\{\emph{rank}(A):A\in H(G)\}$. Then, the following are equivalent:

\begin{enumerate}
\item $\emph{hmr}(G)\leq 2$.
\item $G^c$ has the form $(K_{s_1}\dot\cup\ldots\dot\cup K_{s_t}\dot\cup K_{p_1,q_1}\dot\cup\ldots\dot\cup K_{p_k,q_k})\lor K_r$ where $t$, $s_1,\ldots,s_t$, $k$, $p_1,q_1,\ldots,p_k,q_k$, $r$ are nonnegative integers and $p_i+q_i>0$ for $1\leq i\leq k$.
\item $G$ is $(P_2\dot\cup P_3$, ``fish", $P_2\dot\cup P_2\dot\cup P_2$, ``dart", $P_4)$-free.
\end{enumerate}
\end{theorem}

The proof of Theorem \ref{thm_hmr2}, and the relation between $\text{hmr}(G)$ and $Z(G)$ used in the proof of Theorem \ref{thm_zfn2}, are obtained primarily through linear algebraic techniques. In contrast, in this section, we will develop and use predominantly combinatorial and graph theoretic techniques to derive a characterization of graphs satisfying $Z_c(G)=n-2$.

From Theorem \ref{thm_zfn2}, we can easily derive the following characterization of graphs whose zero forcing number equals $n-2$; this characterization will be used in the sequel.

\begin{corollary}
\label{cor_zfn2}
$Z(G)=n-2$ if and only if $G$ satisfies the following conditions:
\begin{enumerate}
\item $G$ does not contain any of the graphs in Figure \ref{fig_zfn2} as induced subgraphs,
\item $G\not\simeq\dot\bigcup_{i=1}^n K_1$,
\item $G\not\simeq\left(\dot\bigcup_{i=1}^tK_1\right)\dot\cup K_{n-t}$ for $n\geq 2$ and $0\leq t\leq n-2$.
\end{enumerate}
\end{corollary}
\proof
Observe that the second condition in the statement of Corollary \ref{cor_zfn2} is satisfied if and only if $Z(G)=n$, and the third condition is satisfied if and only if $Z(G)=n-1$.
\qed
\vspace{9pt}
\noindent The following is a novel concept in the study of forcing sets, and will be useful in proving a technical lemma. We remark that further study of this restriction of connected forcing (and analogously of zero forcing) would be interesting in its own right.

\begin{definition}
\label{restrained_forcing}
For any $S\subset V$, let $Z_c(G;S)$ be the cardinality of the minimum connected forcing set of $G$ which contains $S$. 
\end{definition}
For example, let $G=(\{1,2,3,4,5\},\{\{1,2\},\{2,3\},\{3,4\},\{4,5\}\})$ be a path on 5 vertices. Then $Z_c(G;\{1\})=1$, $Z_c(G;\{2,3\})=2$, and $Z_c(G;\{1,5\})=5$. 

\begin{lemma}
\label{non_clique_lemma}
Let $G$ be a biconnected graph different from $K_n$. Then for any $v\in V$, $Z_c(G;\{v\})\leq n-2$.
\end{lemma}
\proof
Note that since $G$ is biconnected and not complete, it must have at least 4 vertices. Let $v^*$ be an arbitrary vertex of $G$ and suppose for contradiction that $Z_c(G;\{v^*\})=n-1$. 

Suppose first that some $\{v_1,v_2\}\subset V\backslash\{v^*\}$ forms a separating set of $G$. Let $u$ be a vertex which is not a cut vertex of $G-v_1$ and which belongs to a component of $G-\{v_1,v_2\}$ that does not contain $v^*$ (it is easy to see that such a vertex exists). We claim that $R=V\backslash\{v_1,u\}$ is a connected forcing set of $G$. To see why, note first that by construction $R$ is connected. Moreover, some colored neighbor of $v_1$ in the component of $G-\{v_1,v_2\}$ containing $v^*$ can force $v_1$ in the first time step; then, any neighbor of $u$ can force $u$. Thus, $G$ cannot have a separating set of size 2. 

Let $v$ be any vertex in $V\backslash\{v^*\}$ and suppose there is a vertex $u\in V$ which is not adjacent to $v$; let $w$ be a neighbor of $u$ different from $v^*$ (which exists since $G$ is biconnected). Then, $V\backslash \{v,w\}$ is a connected forcing set of $G$, since $u$ can force $w$ in the first time step, and then $v$ can be forced by any of its neighbors; moreover, since $G$ has no separating sets of size 2, this set is connected. However, since we assumed that $Z_c(G;\{v^*\})=n-1$, it follows that every $v\in V\backslash\{v^*\}$ is adjacent to every vertex in $V$. This implies that $G$ is a complete graph, a contradiction.
\qed
\vspace{9pt}
\noindent The following definition is a generalization of Definition \ref{def_pendant_tree}.
\begin{definition}
A \emph{pendant tree attached to vertex} $v$ in graph $G=(V,E)$ is a set $T\subset V$ composed of the vertices of the connected components of $G-v$ which are trees and which have a single vertex adjacent to $v$ in $G$. 
\end{definition}

\noindent Finally, the following two results will be used in the sequel.
\begin{theorem}
\label{tree_thm}
\emph{\cite{brimkov}} Let $G$ be a tree different from a path; then $M(G)$ is a minimum connected forcing set of $G$.
\end{theorem}
\begin{proposition}\emph{\cite{brimkov2}}
\label{block_prop}
Let $G$ be a connected graph different from a path and $B$ be a block of $G$ which is not a cut edge of a pendant path of $G$. Then every connected forcing set of $G$ contains at least $\delta(G[B])$ vertices of $B$.
\end{proposition}
We now prove the main result of this section.
\begin{theorem}
\label{thm_cfn2}
$Z_c(G)=n-2$ if and only if $G$ belongs to the family of graphs described in Figures \ref{fig_cfn2case1}--\ref{fig_cfn2case6}.
\end{theorem}

\proof
Let $G=(V,E)$ be a graph with $Z_c(G)=n-2$. If $G$ does not have a separating set, then $G$ is a complete graph, and $Z_c(G)=n-1$. Note also that $G$ is connected; thus, $\kappa(G)\geq 1$. We will consider several cases based on the vertex connectivity of $G$, starting with $\kappa(G)=1$. 
We will say $v$ is a \emph{feasible} vertex if $v$ is part of exactly one nontrivial block of $G$ and if every trivial block adjacent to $v$ is part of a pendant tree.
If $v$ is a feasible vertex, define $\ell(v)$ to be $v$ if $v$ is not a cut vertex, and otherwise to be some leaf of $G$ in the pendant tree attached to $v$. Note that for any feasible vertex $v$, deleting $\ell(v)$ does not disconnect $G$.

\begin{claim}
If $G$ is a graph with $\kappa(G)=1$ and if $G$ has three or more nontrivial blocks, then $Z_c(G)\leq n-3$.
\end{claim}

\proof
From the structure of $G$ it follows that $G$ has two nontrivial blocks $B_1$ and $B_2$ with feasible vertices $u_1,v_1$ in $B_1$ and feasible vertices $u_2,v_2$ in $B_2$, and a nontrivial block $B_3$ with a feasible vertex $v_3$. We claim that $V\backslash \{\ell(v_1),\ell(v_2),\ell(v_3)\}$ is a connected forcing set. To see why, note that $\ell(v_1)$ and $\ell(v_2)$ each have a neighbor which is not adjacent to another vertex in $ \{\ell(v_1),\ell(v_2),\ell(v_3)\}$; therefore, $\ell(v_1)$ and $\ell(v_2)$ can be forced in the first time step, and then any neighbor of $\ell(v_3)$ can force $\ell(v_3)$. Thus, $Z_c(G)\leq n-3$.
\qed

\begin{claim}
Let $G$ be a graph with $Z_c(G)=n-2$, $\kappa(G)=1$, and two nontrivial blocks. Then, $G$ belongs to the family of graphs described in Figure \ref{fig_cfn2case1}.
\begin{figure}[ht!]
\begin{center}
\includegraphics[scale=0.55]{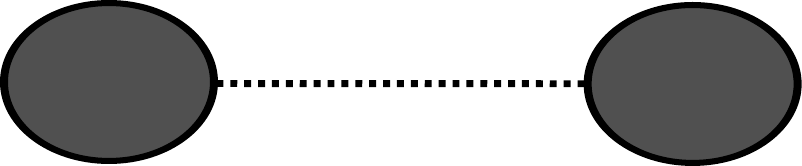}
\caption{Shaded ovals represent cliques, each of size at least 3; dotted line represents a path of possibly zero length.}
\label{fig_cfn2case1}
\end{center}
\end{figure}
\end{claim}
\proof 
Let $B_1$ and $B_2$ be the nontrivial blocks of $G$.  Suppose first that $G$ also has at least one trivial outer block. It is easy to see that there are at least two feasible vertices $u_1,v_1$ in $B_1$ and at least two feasible vertices $u_2,v_2$ in $B_2$. Let $v_3$ be a leaf vertex of some pendant tree of $G$, which, without loss of generality, does not coincide with $\ell(v_1)$ and $\ell(v_2)$ (although it may coincide with $\ell(u_1)$ or $\ell(u_2)$). We claim that $V\backslash \{\ell(v_1),\ell(v_2),v_3\}$ is a connected forcing set. To see why, note that at least one of $\ell(v_1)$ and $\ell(v_2)$, say $\ell(v_1)$, has a neighbor which is not adjacent to another vertex in $\{\ell(v_1),\ell(v_2),v_3\}$. Therefore $\ell(v_1)$ can be forced in the first time step; then, any neighbor of $\ell(v_2)$ can force $\ell(v_2)$, and then the neighbor of $v_3$ can force $v_3$. Thus, $Z_c(G)\leq n-3$, a contradiction.

Now suppose $G$ has no trivial outer blocks, and that at least one of $B_1$ and $B_2$, say $B_1$, is not a clique. Let $v$ be the cut vertex of $B_1$ and $x$ be a non-cut vertex in $B_2$. By Lemma \ref{non_clique_lemma}, $Z_c(G[B_1];\{v\})\leq |B_1|-2$, so there are two vertices $u$ and $w$ in $B_1$ such that $V\backslash \{u,w\}$ is a connected forcing set of $G$. Moreover, some non-cut neighbor of $x$ in $B_2$ can force $x$ in the first time step; thus $V\backslash \{u,w,x\}$ is a connected forcing set of $G$, a contradiction.

Finally, suppose $G$ has no trivial outer blocks, and that both $B_1$ and $B_2$ are cliques. By Proposition \ref{block_prop} and Lemma \ref{MR_lemma}, the set excluding one non-cut vertex from each of $B_1$ and $B_2$ is a minimum connected forcing set of $G$. This is the case illustrated in Figure \ref{fig_cfn2case1}.
\qed

\begin{claim}
\label{tree_claim}
Let $G$ be a graph with $Z_c(G)=n-2$, $\kappa(G)=1$, a single nontrivial block $B$, and non-cut vertex $x\in B$. Then any pendant tree $T$ of $G$ is either composed of one or more leaves attached to a vertex of $B$, or of two or more leaves joined to a vertex of $B$ by a path.
\end{claim}
\proof
Let $T$ be a pendant tree of $G$ attached to some vertex $v\in B$. If $T$ has two leaves $\ell_1$ and $\ell_2$ which are not adjacent to the same vertex, then $V\backslash\{\ell_1,\ell_2,x\}$ is a connected forcing set; thus, all leaves of $T$ are adjacent to the same vertex. $T$ also cannot be a pendant path of length more than 1, since then $V\backslash\{\ell,w,x\}$ is a connected forcing set, where $\ell$ is the leaf of the pendant path and $w$ is the neighbor of $\ell$. Thus, $T$ is composed of one or more leaves attached to $v$, or of two or more leaves joined to $v$ by a path.
\qed

\begin{claim}
Let $G$ be a graph with $Z_c(G)=n-2$, $\kappa(G)=1$, and a single nontrivial block, which is either an inner block or an outer block and a clique. Then, $G$ belongs to the family of graphs described in Figure \ref{fig_cfn2case2}.
\begin{figure}[ht!]
\begin{center}
\includegraphics[scale=0.35]{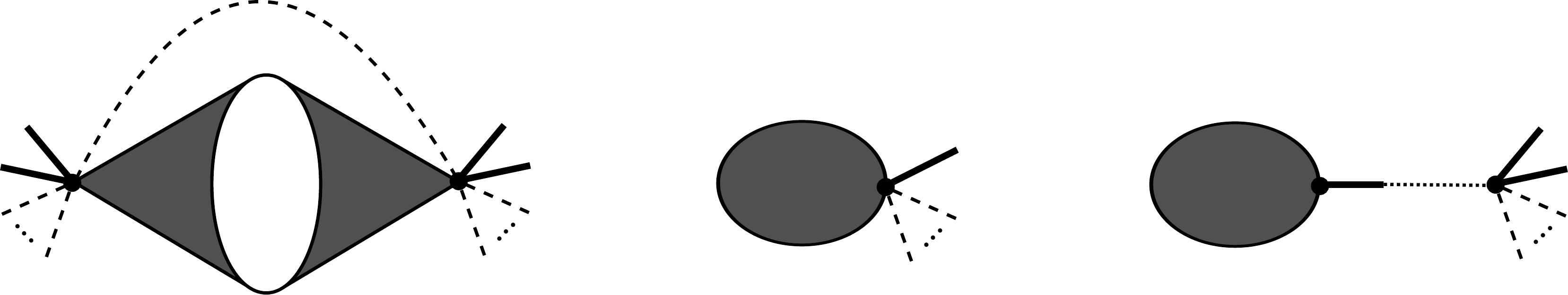}
\caption{White oval represents an independent set of size at least 1, shaded ovals represent cliques of size at least 3. Shaded regions represent all possible edges being present. Dotted line represents a path of possibly zero length, thick lines represent mandatory single edges, dashed straight lines represent an arbitrary number (possibly zero) of single edges. If white oval consists of a single vertex, dashed curved line represents a mandatory single edge; otherwise, it represents  a possibly non-existent single edge.}
\label{fig_cfn2case2}
\end{center}
\end{figure}
\end{claim}

\proof
Let $B$ be the nontrivial block of $G$ and suppose first that $B$ is an inner block. If $B$ has at least 3 cut vertices $v_1$, $v_2$, and $v_3$ (which are by definition feasible vertices), then $V\backslash\{\ell(v_1),\ell(v_2),\ell(v_3)\}$ is a connected forcing set, a contradiction. Thus $B$ has 2 cut vertices $v_1$ and $v_2$. Let $T_1$ and $T_2$ be the pendant trees attached to $v_1$ and $v_2$, respectively, and let $x$ be some non-cut vertex of $B$. By Claim \ref{tree_claim}, for $i\in\{1,2\}$, $T_i$ is composed of one or more leaves attached to $v_i$, or of two or more leaves joined to $v_i$ by a path.

If at least one of $T_1$ and $T_2$, say $T_1$, is composed of two or more leaves joined to $v_1$ by a path, let $\ell_1$ be a leaf in $T_1$, $\ell_2$ be a leaf in $T_2$, and $x$ be a non-cut vertex of $B$. Then $V\backslash\{\ell_1,\ell_2,x\}$ is a connected forcing set, since $\ell_1$ can be forced in the first time step by its neighbor, then any neighbor of $x$ (possibly except $v_2$) can force $x$, and then $\ell_2$ can be forced by its neighbor; this is a contradiction. 

If at least one of $T_1$ and $T_2$, say $T_1$, consists of a single leaf $\ell_1$, and $\ell_2$ is a leaf in $T_2$, then $V\backslash\{\ell_1,v_1,\ell_2\}$ is a connected forcing set since some non-cut neighbor of $v_1$ in $B$ (possibly except $v_2$) can force $v_1$ in the first time step, and then $\ell_1$ and $\ell_2$ can be forced by their neighbors; this is a contradiction. 

Thus, $T_1$ and $T_2$ each consist of two or more leaves. Let $\ell_1$ and $\ell_2$ be leaves in $T_1$ and $T_2$, respectively. If $G[B\backslash\{v_1,v_2\}]$ is not an empty graph, then there is an edge between two vertices $x$ and $y$ in $B\backslash\{v_1,v_2\}$. Then, $V\backslash\{\ell_1,\ell_2,x\}$ is a connected forcing set since $y$ can force $x$ in the first time step, and then $\ell_1$ and $\ell_2$ can be forced by $v_1$ and $v_2$ (note that this set is connected since $x$ is not a cut vertex of $G$); this is a contradiction. 

Now, suppose $G[B\backslash\{v_1,v_2\}]$ is an empty graph. Then both $v_1$ and $v_2$ must be adjacent to every vertex in $B\backslash\{v_1,v_2\}$, since otherwise $G[B]$ would not be biconnected; $v_1$ and $v_2$ could also possibly be adjacent to each other, and if $B\backslash\{v_1,v_2\}$ consists of a single vertex, then $v_1$ and $v_2$ must necessarily be adjacent in order for $G[B]$ to be biconnected. Moreover, it is easy to see that $V\backslash\{\ell_1,\ell_2\}$ is a connected forcing set of $G$. This set is also minimum, since by Lemma \ref{MR_lemma}, $v_1$ and $v_2$ are contained in every connected forcing set of $G$, and a set excluding 3 or more vertices of $V\backslash\{v_1,v_2\}$ will exclude at least two neighbors of at least one of $v_1$ and $v_2$, and will therefore not be forcing. This family of graphs is illustrated in Figure \ref{fig_cfn2case2}, left.

Now suppose that $B$ is an outer block and a clique. Let $v$ be the cut vertex of $B$, $T$ be the pendant tree attached to $v$, $\ell$ be a leaf in $T$, and $x$ be some non-cut vertex of $B$. By Claim \ref{tree_claim}, $T$ is either composed of one or more leaves attached to $v$, or of two or more leaves joined to $v$ by a path. In both cases, by Proposition~\ref{block_prop} and Lemma~\ref{MR_lemma}, $V\backslash\{x,\ell\}$ is a minimum connected forcing set of $G$. These two cases are illustrated in Figure \ref{fig_cfn2case2} middle and right, respectively.
\qed

\begin{claim}
Let $G$ be a graph with $Z_c(G)=n-2$, $\kappa(G)=1$, and a single nontrivial block, which is an outer block and not a clique. Then, $G$ belongs to the family of graphs described in Figure \ref{fig_cfn2case3}.
\begin{figure}[ht!]
\begin{center}
\includegraphics[scale=0.40]{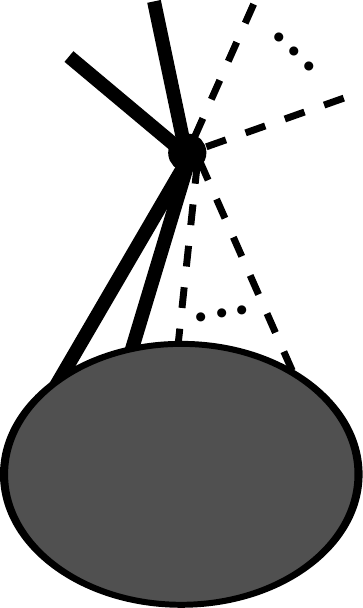}
\caption{Shaded region represents a clique of size at least 3; thick lines represent mandatory single edges, dashed lines represent an arbitrary number (possibly zero) of single edges, such that the vertex outside the shaded oval is not adjacent to every vertex inside the shaded oval.}
\label{fig_cfn2case3}
\end{center}
\end{figure}
\end{claim}

\proof
Let $B$ be the nontrivial block of $G$ and $v$ be the cut vertex of $B$. By Claim \ref{tree_claim}, the pendant tree $T$ attached to $v$ must either be composed of one or more leaves attached to $v$, or two or more leaves joined to $v$ by a path. If $T$ consists of two or more leaves joined to $v$ by a path, then by Lemma \ref{non_clique_lemma}, $Z_c(G[B];\{v\})\leq |B|-2$, so there are two vertices $x$ and $y$ in $B$ such that $V\backslash \{x,y\}$ is a connected forcing set. Moreover, a leaf $\ell$ in $T$ can be forced by its neighbor in the first time step; it follows that $V\backslash \{x,y,\ell\}$ is a connected forcing set of $G$, a contradiction. 



Thus, $T$ consists of one or more leaves attached to $v$. Let $\ell$ be one of these leaves. Suppose $G[B]-v$ has no separating set; then $G[B]-v$ is a clique. Since $G[B]$ is not a clique, there must be some vertex $x\in B\backslash\{v\}$ which is not adjacent to $v$. Let $y\neq x$ be another vertex in $B\backslash\{v\}$. If $T$ consists of a single leaf, then $V\backslash\{\ell,v,y\}$ is a connected forcing set of $G$, since $x$ can force $y$ in the first time step, then any neighbor of $v$ in $B\backslash\{v\}$ can force $v$, and then $v$ can force $\ell$. If $T$ contains two or more leaves, then $V\backslash\{\ell,x\}$ is a connected forcing set. Moreover, this set is minimum, since by Lemma \ref{MR_lemma}, every connected forcing set contains $v$ and all-but-one leaves attached to $v$, and if a set excludes two or more vertices from $B\backslash\{v\}$, then any colored neighbor of these vertices would always have at least two uncolored neighbors. This family of graphs is illustrated in Figure \ref{fig_cfn2case3}.

Now suppose $G[B]-v$ does have a separating set. Note that since $G[B]$ is biconnected, $\kappa(G[B]-v)\geq 1$. If $\kappa(G[B]-v)=1$, let $u$ be a vertex such that $\{u,v\}$ is a separating set of $G$ and let $x\in B$ be a non-cut vertex of $G-u$; let $C$ be the component of $G-\{u,v\}$ containing $x$. Then $V\backslash\{\ell,u,x\}$ is a connected forcing set, since $u$ can be forced by some neighbor of $u$ in a component of $G-\{u,v\}$ other than $C$ in the first time step, then $x$ can be forced by any of its neighbors except $v$, and then $v$ can force $\ell$. Thus, $Z_c(G)\leq n-3$, a contradiction.

If $\kappa(G[B]-v)\geq 2$ and $d_{G[B]}(v)=2$, let $u$ and $w$ be the neighbors of $v$ in $B$. Suppose first that there is some vertex $x\in B\backslash\{u,v,w\}$ such that at least one of $\{u,x\}$ and $\{w,x\}$, say $\{u,x\}$, is a separating set of $G$. Let $C$ be a component of $G-\{u,x\}$ which does not contain $v$. Let $y$ be a non-cut vertex of $G-x$ in $C$. Then $V\backslash \{x,y,\ell\}$ is a forcing set of $G$, since some neighbor of $x$ (except $v$) in a component of $G-\{u,x\}$ other than $C$ can force $x$ in the first time step, then any neighbor of $y$ can force $y$, and then $v$ can force $\ell$. This set is also connected since $y$ is a non-cut vertex of $G[B]-x$, which is connected. Thus, $Z_c(G)\leq n-3$, a contradiction.

%
%
%
%
%

Now suppose that for any $x\in B\backslash\{u,v,w\}$, neither $\{u,x\}$ nor $\{w,x\}$ is a separating set of $G$. If $u$ is not adjacent to $w$, let $y$ be any neighbor of $u$ in $B\backslash\{u,v,w\}$. Then $V\backslash\{\ell,w,y\}$ is a connected forcing set, since $u$ can force $y$ in the first time step, then any neighbor of $w$ (except $v$) can force $w$, and then $v$ can force $\ell$. Note that this set is connected, since by assumption, $\{w,y\}$ is not a separating set of $G$. If $u$ is adjacent to $v$, suppose there is a vertex $y\in B\backslash\{u,v,w\}$ which is not adjacent to at least one of $u$ and $w$, say $y\not\sim u$. Then $V\backslash\{\ell,w,y\}$ is a connected forcing set, since $u$ can force $w$ in the first time step, then any neighbor of $y$ can force $y$, and then $v$ can force $\ell$. Now suppose every vertex in $B\backslash\{u,v,w\}$ is adjacent to both $u$ and $w$. Since $G[B]-v$ is not a clique, there must be some vertices $x$ and $y$ in $B\backslash\{u,v,w\}$ which are not adjacent to each other. Then $V\backslash\{\ell,w,y\}$ is a connected forcing set, since $x$ can force $w$ in the first time step, then any neighbor of $y$ can force $y$, and then $v$ can force $\ell$. In all these cases, it follows that $Z_c(G)\leq n-3$, a contradiction.

If $\kappa(G[B]-v)=2$ and $d_{G[B]}(v)\geq 3$, let $\{u,w\}$ be a separating set of $G[B]-v$, and let $x$ be a non-cut vertex of $G[B]-\{u,v\}$ in some component $C$ of $G[B]-\{u,v,w\}$. Then $V\backslash \{u,x,\ell\}$ is a forcing set of $G$, since any neighbor of $u$ in a component of $G[B]-\{u,v,w\}$ different from $C$ can force $u$ in the first time step, then any neighbor of $x$ (except $v$) can force $x$, and then $v$ can force $\ell$. This set is also connected: $G[B]-\{u,v\}$ is connected since $\kappa(G[B]-v)=2$, $G[B]-\{u,v,x\}$ is connected since $x$ is a non-cut vertex of $G[B]-\{u,v\}$, $G[B]-\{u,x\}$ is connected since $d_{G[B]}(v)\geq 3$ and hence $v$'s neighbors in $B$ cannot be only $u$ and $x$, and $G-\{u,x,\ell\}$ is connected since $\ell$ is a leaf. Thus, $Z_c(G)\leq n-3$, a contradiction.

If $\kappa(G[B]-v)\geq 3$ and $d_{G[B]}(v)\geq 3$, let $x$ and $y$ be two non-adjacent vertices in $B\backslash\{v\}$, and let $z\in B\backslash\{v\}$ be a neighbor of $y$. Then $V\backslash\{\ell,x,z\}$ is a forcing set, since $y$ can force $z$ in the first time step, then any neighbor of $x$ except $v$ can force $x$, and then $v$ can force $\ell$. This set is also connected, since $\kappa(G[B]-v)\geq 3$ and since $d_{G[B]}(v)\geq 3$. Thus, $Z_c(G)\leq n-3$, a contradiction.
\qed

\begin{claim}
Let $G$ be a graph with $Z_c(G)=n-2$, $\kappa(G)=1$, and no nontrivial blocks. Then, $G$ is one of the graphs described in Figure \ref{fig_cfn2case4}.
\begin{figure}[ht!]
\begin{center}
\includegraphics[scale=0.35]{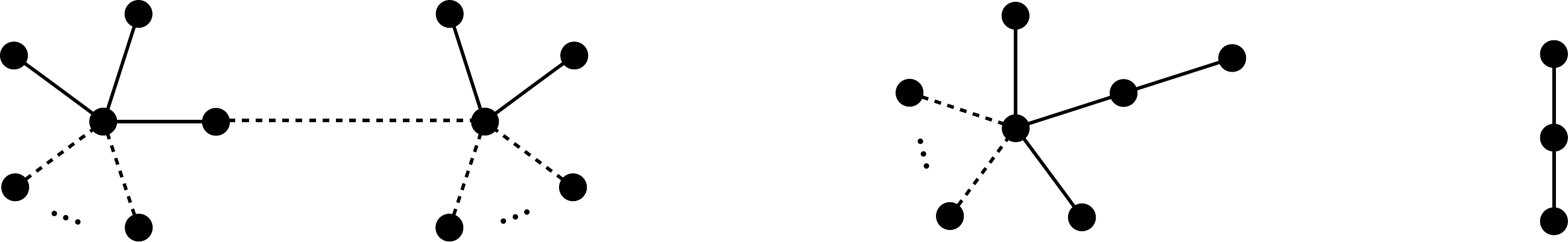}
\caption{Left: two stars each with at least 2 leaves joined by a path of length at least 1. Middle: Pendant attached to a leaf of a star with at least 3 leaves. Right: $P_3$.}
\label{fig_cfn2case4}
\end{center}
\end{figure}
\end{claim}

\proof
Since $G$ has only trivial blocks, $G$ is a tree; thus, by Theorem \ref{tree_thm}, $Z_c(G)=R_2\cup R_3\cup \mathcal{L}$. If $G$ has three or more vertices in $R_3$, then there are at least 3 vertices not in $\mathcal{L}$, so $Z_c(G)\leq n-3$, a contradiction.

If $G$ has two vertices $u$ and $v$ in $R_3$, then all other vertices must belong to a path connecting $u$ and $v$, or to pendant paths attached to $u$ or $v$. By a similar argument as in Claim \ref{tree_claim}, the length of any pendant path attached to $u$ or $v$ must be 1. By Theorem \ref{tree_thm}, this condition is also sufficient to guarantee that $Z_c(G)=n-2$. This is the family of graphs illustrated in Figure \ref{fig_cfn2case4}, left.

If $G$ has one vertex $v$ in $R_3$, then all other vertices must belong to pendant paths attached to $v$. If all pendant paths have length 1, then $G$ is a star and $Z_c(G)=n-1$. If more than one pendant path has length greater than 1, or if any pendant path has length greater than 2, by a similar argument as in Claim \ref{tree_claim}, it follows that $Z_c(G)\leq n-3$. Thus, one pendant path must have length 2, and all other pendant paths must have length 1. This is the family of graphs illustrated in Figure \ref{fig_cfn2case4}, middle.

If $G$ has no vertices in $R_3$, then $G$ is a path, and $Z_c(G)=n-2$ if and only if $G\simeq P_3$. This is the graph illustrated in Figure \ref{fig_cfn2case4}, right.
\qed

\begin{claim}
If $G$ is a graph with $\kappa(G)\geq 2$ and $S$ is a minimum separating set of $G$ such that $G-S$ has three or more components, at least one of which is nontrivial, then $Z_c(G)\leq n-3$. 
\end{claim}

\proof
Let $B_1,\ldots,B_k$ be the components of $G-S$, $k\geq 3$; let $s_1$ and $s_2$ be vertices in $S$. Since $S$ is minimum, each vertex of $S$ is connected to at least one vertex of every component of $G-S$. Without loss of generality, let $B_1$ be a nontrivial component of $G-S$. Note that since $G[B_1]$ is connected and nontrivial, it has at least two non-cut vertices. If $s_1$ is adjacent to exactly one vertex of $B_1$, let $x_1$ be a non-cut vertex of $G[B_1]$ different from the neighbor of $s_1$ in $B_1$; otherwise, if $s_1$ is adjacent to two or more vertices of $B_1$, let $x_1$ be an arbitrary non-cut vertex of $G[B_1]$. If $B_2$ is a trivial block, let $x_2$ be the vertex of $B_2$; if $B_2$ is nontrivial and if $s_1$ is adjacent to exactly one vertex of $B_2$, let $x_2$ be a non-cut vertex of $G[B_2]$ different from the neighbor of $s_1$ in $B_2$, and if $s_1$ is adjacent to two or more vertices of $B_2$, let $x_2$ be an arbitrary non-cut vertex of $G[B_2]$. In every case, $V\backslash\{s_2,x_1,x_2\}$ is a forcing set, since any neighbor of $s_2$ in $B_3$ can force $s_2$ in the first time step, then any neighbor of $x_1$ in $B_1$ can force $x_1$, and then any neighbor of $x_2$ can force $x_2$. This set is also connected, since each of the graphs  $G[B_1]-x_1,G[B_2]-x_2,G[B_3],\ldots,G[B_k]$ is connected, $s_1$ is connected to each of these graphs, and all other vertices of $S\backslash\{s_2\}$ are connected to some vertices in $G[B_3],\ldots,G[B_k]$. Thus, $Z_c(G)\leq n-3$.
\qed

\begin{claim}
\label{claim_trivial_comp}
Let $G$ be a graph with $Z_c(G)=n-2$ and let $S$ be a minimum separating set of $G$ such that $G-S$ has only trivial components. Then every trivial component of $G-S$ must be adjacent to every vertex in $S$; moreover, any connected forcing set of $G$ excludes at most one trivial component of $G-S$.
\end{claim}
\proof
Let $v$ be a vertex that is a trivial component of $G-S$. If $v$ is not adjacent to some vertex $u\in S$, then $S\backslash \{u\}$ would be a smaller separating set of $G$ than $S$. Let $R$ be a connected forcing set of $G$ and suppose $R$ excludes two vertices $v_1$ and $v_2$ which are trivial components of $G-S$. Since $v_1$ and $v_2$ are only adjacent to vertices in $S$, and since every vertex in $S$ has at least two uncolored neighbors (namely $v_1$ and $v_2$), no vertex in $S$ would be able to force $v_1$ and $v_2$. Thus, any connected forcing set can exclude at most one trivial component of $G-S$.
\qed

\begin{claim}
\label{claim_all_trivial}
Let $G$ be a graph with $Z_c(G)=n-2$, $\kappa(G)\geq 2$, and let $S$ be a minimum separating set of $G$ such that $G-S$ has only trivial components. Then $G$ is one of the graphs described in Figure \ref{fig_cfn2case5}.
\begin{figure}[ht!]
\begin{center}
\includegraphics[scale=0.25]{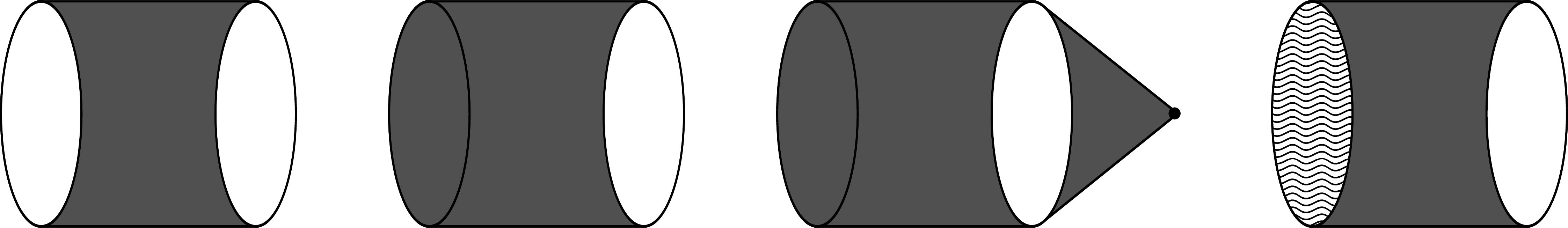}
\caption{Ovals represent sets of vertices, each of size at least two. Shaded regions represent all possible edges being present within a set of vertices or between sets of vertices; white regions represent no edges being present. Region with wave pattern represents a set of vertices which induces a graph $H$ which has no isolated vertices and has zero forcing number $|V(H)|-2$.}
\label{fig_cfn2case5}
\end{center}
\end{figure}
\end{claim}
\proof
Let $C$ be the set of vertices which are trivial components of $G-S$. By Claim \ref{claim_trivial_comp}, every vertex in $C$ is adjacent to every vertex in $S$. Suppose for contradiction that $Z(G[S])<|S|-2$, and let $Z$ be a minimum zero forcing set of $G[S]$. Then $Z\cup C$ is a connected forcing set of $G$, since any vertex in $Z$ which forces a vertex in $G[S]$ can also force the same vertex in $G$. Thus, $Z_c(G)\leq |Z|+|C|<|S|-2+|C|=n-2$, a contradiction. Thus, $Z(G[S])\geq |S|-2$.

If $Z(G[S])=|S|$, then $G[S]$ is an empty graph, and hence $G$ is a complete bipartite graph with parts $C$ and $S$. Then, any set containing all-but-one vertices of $C$ and all-but-one vertices of $S$ is connected and forcing (note that $|C|\geq 2$ and $|S|\geq 2$). This set is also minimum, since a set excluding more than one vertex from one (or both) of $C$ and $S$ is not forcing. This family of graphs is illustrated in Figure \ref{fig_cfn2case5}, left.

If $Z(G[S])=|S|-1$, then $G[S]$ is the disjoint union of a nontrivial clique and zero or more isolated vertices. If $G[S]$ has at most one isolated vertex, then any set containing all-but-one vertices of $C$ and all-but-one vertices of $S$ is connected and forcing. This set is also minimum, since by Claim \ref{claim_trivial_comp}, a connected forcing set $R$ can exclude at most one vertex of $C$; if $R$ excludes one vertex of $C$, then it cannot exclude two or more vertices of $S$, since then no vertex will be able to force them. Similarly, if $R$ contains all vertices of $C$, then it cannot exclude three or more vertices of $S$, since then at least two of them will belong to the nontrivial clique in $G[S]$, and no vertex will be able to force them. Thus $Z_c(G)=n-2$; this family of graphs is illustrated in Figure \ref{fig_cfn2case5}, middle-left and middle-right.

If $G[S]$ is the disjoint union of a nontrivial clique and two or more isolated vertices, then let $u$ and $x$ be isolated vertices in $G[S]$, $v$ and $y$ be vertices in the nontrivial clique of $G[S]$, and $w$ be a vertex in $C$. Then $V\backslash\{u,v,w\}$ is a connected forcing set, since $x$ can force $w$ in the first time step, then $y$ can force $v$, and then $w$ can force $u$. Thus $Z_c(G)\leq n-3$, a contradiction.

Finally, if $Z(G[S])=|S|-2$, then $G[S]$ is one of the graphs in Corollary \ref{cor_zfn2}. 
Let $Z$ be an arbitrary minimum zero forcing set of $G[S]$, let $\{z_1,z_2\}=S\backslash Z$, and let $x$ be a vertex in $C$. If $G[S]$ has an isolated vertex $v$, then $v$ must be contained in $Z$. Then $V\backslash \{z_1,z_2,x\}$ is a connected forcing set, since $v$ can force $x$ in the first time step, and then $z_1$ and $z_2$ can be forced by the same vertices which force them in $G[S]$. Thus, $G[S]$ does not have isolated vertices. Moreover, $V\backslash\{z_1,z_2\}$ is a connected forcing set of $G$, since $z_1$ and $z_2$ can be forced in $G$ by the same vertices which force them in $G[S]$; we claim that this set is also minimum. To see why, suppose there is a connected forcing set $R$ which excludes three or more vertices of $G$. By Claim \ref{claim_trivial_comp}, $R$ can exclude at most one vertex of $C$. If $R$ excludes one vertex $x$ of $C$ and two or more vertices of $S$, then no vertex in $C$ can force another vertex until all-but-one vertices in $S$ are forced (because until then, all vertices in $C$ are adjacent to two or more uncolored vertices in $S$). Thus, the first force must be performed by a vertex $y$ in $S$. This means $y$ has a single uncolored neighbor, which must be $x$. Then, all neighbors of $y$ in $S$ are contained in $R$. Let $R'$ be the set obtained by adding $x$ and all-but-two vertices in $S\backslash R$ to $R$. $R'$ is also connected and forcing, and there is a chronological list of forces where both vertices not in $R'$ are forced by vertices of $S$. Thus $R'\cap S$ is a zero forcing set of $G[S]$ of size $|S|-2$. However, $y$ is a non-isolated vertex in $G[S]$, which is in $R'\cap S$ and all of whose neighbors are in $R'\cap S$. Therefore, $R'\cap S\backslash\{z\}$ is also a zero forcing set of $G[S]$, where $z$ is a neighbor of $y$ in $S$; this contradicts the assumption that $Z(G[S])=|S|-2$. Similarly, if $R$ excludes no vertices of $C$, then it cannot exclude three or more vertices of $S$, since then $S\cap R$ would be a zero forcing set of $G[S]$ of size at most $|S|-3$, a contradiction. Thus, $Z_c(G)=n-2$; this family of graphs is illustrated in Figure \ref{fig_cfn2case5}, right.
\qed

\begin{claim}
\label{claim_full_join2}
Let $G$ be a graph with $Z_c(G)=n-2$ and $\kappa(G)=2$; let $S$ be a minimum separating set of $G$ such that $G-S$ has exactly two components, at least one of which is nontrivial. Then each component of $G-S$ is a clique, and each vertex from each component of $G-S$ is adjacent to every vertex in $S$. 
\end{claim}
\proof
Let $s_1$ and $s_2$ be the vertices of $S$, and let $B_1$ and $B_2$ be the components of $G-S$. Let $\mathcal{I}=\{\{(1,1),(2,1)\},\{(1,1),(2,2)\},\{(1,2),(2,1)\},\{(1,2),(2,2)\}\}$ and $\mathcal{J}=\{\{(1,1),(1,2)\},\{(2,1),(2,2)\}\}$.

Suppose first that there exists a set $I\in \mathcal{I}$ such that for each $(i,j)\in I$, $G[B_i\cup \{s_j\}]$ has no cut vertices. Without loss of generality, let $I=\{(1,1),(2,1)\}$, i.e., suppose $G[B_1\cup \{s_1\}]$ and $G[B_2\cup \{s_1\}]$ have no cut vertices. Suppose also that $s_2$ is not adjacent to some vertex of $B_1\cup B_2$, say $x\in B_2$; then, $B_2$ must be a nontrivial component. Let $y$ be a neighbor of $x$ in $B_2$ and let $v\in B_1$ be a non-cut vertex of $G-s_2$. Then, $V\backslash \{v,s_2,y\}$ is a forcing set, since $x$ can force $y$ in the first time step, then any neighbor of $s_2$ in $B_2$ can force $s_2$, and then any neighbor of $v$ can force $v$. This set is also connected, since $G-s_2$ is connected, $y$ is not a cut vertex of $G[B_2\cup \{s_1\}]$, and $v$ is not a cut vertex of $G[B_1\cup \{s_1\}]$. This contradicts $Z_c(G)= n-2$, so $s_2$ must be adjacent to every vertex in $B_2$. Hence, $G[B_2\cup \{s_2\}]$ has no cut vertices (since $G[B_2]$ is connected), and so by the same argument as above, it follows that $s_1$ is also adjacent to every vertex in $B_2$. Similarly, $s_1$ and $s_2$ are adjacent to every vertex in $B_1$. Now suppose $B_2$ is not a clique; then, $B_2$ must have at least three vertices. Let $x$ and $y$ be two non-adjacent vertices in $B_2$; let $z$ be a neighbor of $x$ in $B_2$, and let $v$ be any vertex in $B_1$. Then, $V\backslash\{y,z,v\}$ is a connected forcing set, since $x$ can force $z$ in the first time step, then any neighbor of $y$ in $B_2$ can force $y$, and then any neighbor of $v$ can force $v$. This set is also connected, since every vertex in $B_1$ and $B_2$ is adjacent to $S$, and every vertex in $S$ is adjacent to $x$. This is a contradiction, so $B_2$ is a clique; similarly, $B_1$ is a clique.

Now suppose that there does not exist a set $I\in \mathcal{I}$ such that for each $(i,j)\in I$, $G[B_i\cup \{s_j\}]$ has no cut vertices. 
Equivalently, there exists a set $J\in \mathcal{J}$ such that for each $(i,j)\in J$, $G[B_i\cup \{s_j\}]$ has cut vertices. Without loss of generality, let $J=\{(2,1),(2,2)\}$, i.e., $G[B_2\cup \{s_1\}]$ and $G[B_2\cup \{s_2\}]$ have cut vertices. Hence, $G[B_2]$ has cut vertices, since $G[B_2]$ is connected. At least one of $s_1$ and $s_2$ must be adjacent to a non-cut vertex of every outer block of $G[B_2]$, since otherwise the cut vertex of such a block would be a cut vertex of $G$. Note also that if $s_1$ or $s_2$, say $s_1$, is adjacent to a non-cut vertex of every outer block of $G[B_2]$, then $G[B_2\cup \{s_1\}]$ would not have any cut vertices. Thus, there is an outer block $D_1$ of $G[B_2]$ such that $s_1$ is adjacent to a non-cut vertex $d_1$ of $D_1$ and $s_2$ is not adjacent to any non-cut vertex of $D_1$, and there is an outer block $D_2$ of $G[B_2]$ such that $s_2$ is adjacent to a non-cut vertex $d_2$ of $D_2$ and $s_1$ is not adjacent to any non-cut vertex of $D_2$.

Suppose $s_1$ is adjacent to a single vertex of $B_2$; this must be the vertex $d_1$ defined above. Let $v\in B_1$ be a non-cut vertex of $G-s_1$. Then, $V\backslash\{v,s_1,d_1\}$ is a forcing set, since any neighbor of $d_1$ in $B_2$ can force $d_1$ in the first time step, then $d_1$ can force $s_1$, and then any neighbor of $v$ can force $v$. This set is also connected, since $G-s_1$ is connected, and $v$ and $d_1$ are non-cut vertices of $G-s_1$. Thus, $Z_c(G)\leq n-3$, a contradiction.

Now suppose $s_1$ is adjacent to two or more vertices of $B_2$. Let $v\in B_1$ be a non-cut vertex of $G-s_2$. Then, $V\backslash\{v,s_2,d_1\}$ is a forcing set, since $d_2$ can force $s_2$ in the first time step, then $d_1$ can be forced by any of its neighbors in $B_2$, and then any neighbor of $v$ can force $v$. This set is also connected, since $G[B_1]-v$ is connected, $G[B_2]-d_1$ is connected, $s_1$ is adjacent to some vertex in $B_1$ other than $v$, and $s_1$ is adjacent to some vertex in $B_2$ other than $d_1$. Thus, $Z_c(G)\leq n-3$, a contradiction. 
\qed

\begin{claim}
\label{claim_full_join3}
Let $G$ be a graph with $Z_c(G)=n-2$ and $\kappa(G)=3$; let $S$ be a minimum separating set of $G$ such that $G-S$ has exactly two components, at least one of which is nontrivial. Then each component of $G-S$ is a clique, and each vertex from each component of $G-S$ is adjacent to every vertex in $S$. 
\end{claim}

\proof
Suppose first that at least one of $B_1$ and $B_2$, say $B_2$, is not a clique; then $B_2$ is a nontrivial component. Suppose also that no two vertices of $B_2$ form a separating set of $G-s_1$. Let $x$ and $y$ be two nonadjacent vertices in $B_2$, and let $z$ be a neighbor of $x$ in $B_2$. Then, $V\backslash\{s_1,y,z\}$ is a forcing set, since any neighbor of $s_1$ in $B_1$ can force $s_1$ in the first time step, then $x$ can force $z$, and then any neighbor of $y$ can force $y$. This set is also connected, since $G-s_1$ is connected, and by assumption $\{y,z\}$ is not a separating set of $G-s_1$. 

Now suppose that two vertices $t_1$ and $t_2$ in $B_2$ form a separating set of $G-s_1$. Let $D$ be a component of $G-\{s_1,t_1,t_2\}$ which does not contain $s_2$ and $s_3$. Note that $s_1$ must be adjacent to some vertex $d$ in $D$, since otherwise $\{t_1,t_2\}$ would be a separating set of $G$. Let $v\in B_1$ be a non-cut vertex of $G-\{s_1,s_2\}$. Then, $V\backslash\{s_1,s_2,v\}$ is a forcing set, since $d$ can force $s_1$ in the first time step, then any neighbor of $s_2$ in $B_2$ can force $s_2$, and then any neighbor of $v$ can force $v$. This set is also connected, since $G-\{s_1,s_2\}$ is connected, and $v$ is a non-cut vertex of $G-\{s_1,s_2\}$. 

In both cases, it follows that $Z_c(G)\leq n-3$, a contradiction; thus, $B_2$ is a clique, and similarly, $B_1$ is a clique. Now suppose that some vertex in $S$, say $s_1$, is not adjacent to some vertex in $B_1$ or $B_2$, say $x\in B_2$; note that $B_2$ must then be a nontrivial component. Let $v$ be any vertex in $B_1$.

If $|B_2|=2$, let $B_2=\{x,y\}$. Then, $s_1$ is adjacent only to $y$, so both $s_2$ and $s_3$ must be adjacent to $x$, since otherwise $x$ will have fewer than three neighbors (contradicting $\kappa(G)=3$). Then, $V\backslash\{s_1,y,v\}$ is a forcing set, since $x$ can force $y$ in the first time step, then any neighbor of $s_1$ in $B_2$ can force $s_1$, and then any neighbor of $v$ can force $v$. This set is also connected, since $s_2$ and $s_3$ are both adjacent to $x$, and if $B_1$ is not a trivial component, then at least one of $s_2$ and $s_3$ is adjacent to a vertex of $B_1$ other than $v$. 

If $|B_2|=3$, let $B_2=\{x,y_1,y_2\}$. Note that any pair of vertices in $B_2$ must collectively have at least two neighbors in $S$, since otherwise their single neighbor and the other vertex in $B_2$ form a separating set of $G$. If $s_1$ is adjacent to a single vertex in $B_2$, let $y$ be that vertex. If $s_1$ is adjacent to both $y_1$ and $y_2$, and if $x$ is adjacent to both $s_2$ and $s_3$, let $y$ be $y_1$. If $s_1$ is adjacent to both $y_1$ and $y_2$, and if $x$ is adjacent to a single vertex $s\in \{s_1,s_2\}$, and if $S\backslash\{s_1,s\}$ has a single neighbor $z\in B_2$, let $y$ be $B_2\backslash\{x,z\}$; if $S\backslash\{s_1,s\}$ has multiple neighbors in $B_2$, let $y$ be $y_1$. In each of these cases, $V\backslash\{s_1,y,v\}$ is a forcing set, since $x$ can force $y$ in the first time step, then any neighbor of $s_1$ in $B_2$ can force $s_1$, and then any neighbor of $v$ can force $v$. This set is also connected, since $G[B_2]-y$ is connected, $G[B_1]-v$ is connected, $s_2$ and $s_3$ are each adjacent to at least one vertex in $G[B_2]-y$ (for each choice of $y$ above), and at least one of $s_2$ and $s_3$ is adjacent to a vertex of $G[B_1]-v$ (if $B_1$ is not a trivial component).

If $|B_2|\geq 4$, let $y$ be a vertex in $B_2$ which is different from $x$, and --- if one or both of $s_2$ or $s_3$ have a single neighbor in $B_2$ --- is different from those neighbors. Then, $V\backslash\{s_1,y,v\}$ is a connected forcing set by the same reasoning as above. 

In all cases, we reach a contradiction, so it follows that each vertex of $B_2$ is adjacent to each vertex of $S$. Similarly, we conclude that each vertex of $B_1$ is adjacent to each vertex of $S$.
\qed

\begin{claim}
\label{claim_full_join4}
Let $G$ be a graph with $Z_c(G)=n-2$ and $\kappa(G)\geq 4$; let $S$ be a minimum separating set of $G$ such that $G-S$ has exactly two components, at least one of which is nontrivial. Then each component of $G-S$ is a clique, and each vertex from each component of $G-S$ is adjacent to every vertex in $S$. 
\end{claim}
\proof
Let $B_1$ and $B_2$ be the components of $G-S$, and suppose for contradiction that some vertex in $B_1$ or $B_2$, say $x\in B_1$ is not adjacent to some vertex in $S$, say $s_1$; note that $B_1$ must then be a nontrivial component. Let $y$ be a neighbor of $x$ in $B_1$ and let $z$ be a vertex in $B_2$. Then, $V\backslash\{y,z,s_1\}$ is a forcing set of $G$, since $x$ can force $y$ in the first time step, then some neighbor of $s_1$ in $B_1$ can force $s_1$, and then any neighbor of $z$ can force $z$. This set is also connected since $\kappa(G)\geq 4$; thus, $Z_c(G)\leq n-3$, a contradiction. Therefore, each vertex from each component of $G-S$ is adjacent to every vertex in $S$.

Now suppose for contradiction that some component of $G-S$, say $B_1$, is not a clique. Note that $B_1$ must then have at least 3 vertices, since if $B_1$ is a trivial component or has two vertices which are connected, then $B_1$ is a clique. Let $x$ and $y$ be vertices in $B_1$ which are not adjacent, and let $z$ be a neighbor of $x$ in $B_1$; let $w$ be a vertex in $B_2$. Then, $V\backslash\{y,z,w\}$ is a forcing set of $G$, since $x$ can force $z$ in the first time step, then some neighbor of $y$ in $B_1$ can force $y$, and then any neighbor of $w$ can force $w$. This set is also connected since $\kappa(G)\geq 4$; thus, $Z_c(G)\leq n-3$, a contradiction. Therefore, each component of $G-S$ is a clique.
\qed

\begin{claim}
Let $G$ be a graph with $Z_c(G)=n-2$, $\kappa(G)\geq 2$ and let $S$ be a minimum separating set of $G$ such that $G-S$ has exactly two components, at least one of which is nontrivial. Then $G$ is one of the graphs described in Figure~\ref{fig_cfn2case6}. 
\begin{figure}[ht!]
\begin{center}
\includegraphics[scale=0.20]{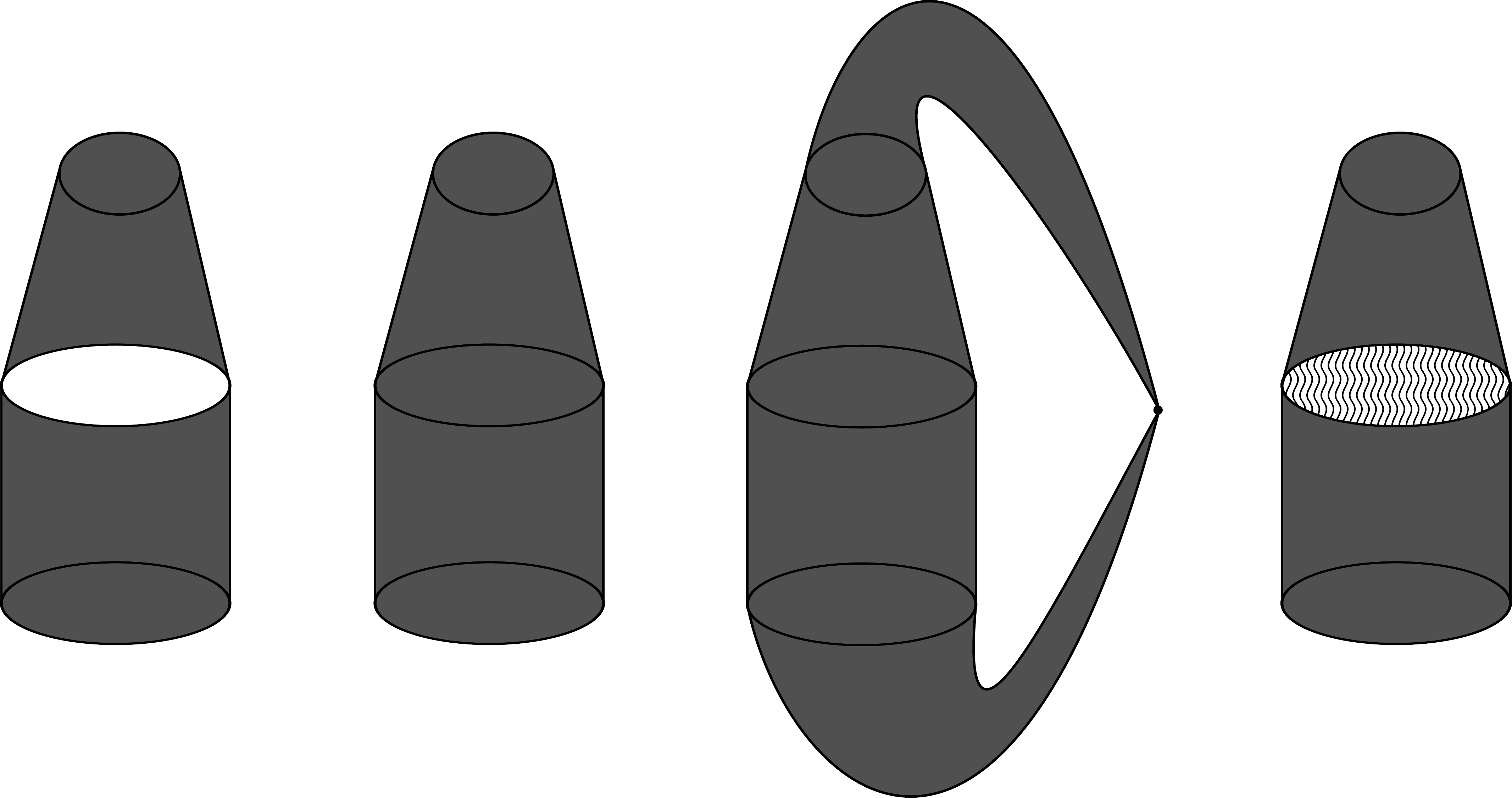}
\caption{Ovals represent sets of vertices. Shaded regions represent all possible edges being present within a set of vertices or between sets of vertices; white regions represent no edges being present. Region with wave pattern represents a set of vertices which induces a graph $H$ which has no isolated vertices and has zero forcing number $|V(H)|-2$. Smaller ovals have at least one vertex, larger ovals have at least two vertices.}
\label{fig_cfn2case6}
\end{center}
\end{figure}
\end{claim}

\proof
Let $B_1$ and $B_2$ be the components of $G-S$, where $B_1$ is a nontrivial component. 
By Claims \ref{claim_full_join2}, \ref{claim_full_join3}, and \ref{claim_full_join4}, $B_1$ and $B_2$ are cliques, and every vertex in $B_1$ and $B_2$ is adjacent to every vertex in $S$. By the same argument as in Claim \ref{claim_all_trivial}, $Z(G[S])\geq |S|-2$.

If $Z(G[S])=|S|$, then $G[S]$ is an empty graph, and any set excluding a single vertex from $S$ and a single vertex from $B_1$ is connected and forcing (note that $|S|\geq 2$). This set is also minimum, since if $R$ is a set which excludes two or more vertices from $B_1$, $S$, or $B_2$, or excludes one vertex from each of $B_1$, $S$, and $B_2$, then every vertex in $R$ will have at least two neighbors not in $R$, and hence $R$ will not be forcing. This family of graphs is illustrated in Figure \ref{fig_cfn2case6}, left.

If $Z(G[S])=|S|-1$, then $G[S]$ is the disjoint union of a clique and zero or more isolated vertices. If $G[S]$ has at most one isolated vertex, then any set excluding a single vertex from $S$ and a single vertex from $B_1$ is connected and forcing. This set is also minimum since if $R$ is a set which excludes two or more vertices from $B_1$, $S$, or $B_2$, or excludes one vertex from each of $B_1$, $S$, and $B_2$, then every vertex in $R$ will have at least two neighbors not in $R$, and hence $R$ will not be forcing. This family of graphs is illustrated in Figure \ref{fig_cfn2case6}, middle-left and middle-right. 

If $G[S]$ is the disjoint union of a clique and two or more isolated vertices, then let $x_1$ and $x_2$ be isolated vertices in $G[S]$, $v_1$ and $v_2$ be vertices in the nontrivial clique of $G[S]$, $u_1$ be a vertex in $B_1$. Then $V\backslash\{u_1,v_1,w_1\}$ is a connected forcing set, since $x_2$ can force $u_1$ in the first time step, then $v_2$ can force $v_1$, and then any neighbor of $x_1$ can force $x_1$. Thus $Z_c(G)\leq n-3$, a contradiction.

Finally, if $Z(G[S])=|S|-2$, then $G[S]$ is one of the graphs in Corollary \ref{cor_zfn2}. 
Let $Z$ be an arbitrary minimum zero forcing set of $G[S]$, and let $\{z_1,z_2\}=S\backslash Z$. By a similar argument as in Claim \ref{claim_all_trivial}, $G[S]$ does not have isolated vertices; moreover, $V\backslash\{z_1,z_2\}$ is a connected forcing set of $G$. We claim that this set is also minimum; to see why, suppose there is a connected forcing set $R$ which excludes three or more vertices of $G$. If $R$ excludes three or more vertices of $B_1\cup B_2$, then two of them are in the same clique component of $G-S$, and can therefore not be forced by any of their neighbors. For the same reason, if $R$ excludes two vertices of $B_1\cup B_2$, then one of these vertices must be in $B_1$ and the other must be in $B_2$; however, if $R$ also excludes one or more vertex of $S$, then every vertex of $G$ will have at least two uncolored neighbors, and no forcing will be possible. By a similar argument as in Claim \ref{claim_all_trivial}, we also reach a contradiction if $R$ excludes one vertex of $B_1\cup B_2$ and two or more vertices of $S$, or if $R$ excludes no vertices of $B_1\cup B_2$ and three or more vertices of $S$. Thus, $Z_c(G)=n-2$; this family of graphs is illustrated in Figure \ref{fig_cfn2case6}, right.
\qed
\vspace{9pt}
Since each of the graphs described in Figures \ref{fig_cfn2case1}--\ref{fig_cfn2case6} has connected forcing number $n-2$, this concludes the proof of Theorem \ref{thm_cfn2}.
\qed
\vspace{9pt}
The statement of Theorem \ref{thm_cfn2} can be rewritten in a similar format as the statement of Theorem \ref{thm_hmr2}; however, we chose to express our results using explicit diagrams in order to make it easier to visualize the structure of the graphs in question. Due to the constant number of equivalence classes of vertices in each of the graphs in Figures \ref{fig_cfn2case1}--\ref{fig_cfn2case6} (or in their complements, according to Theorem \ref{thm_hmr2}), it is readily verifiable that a graph in this family is efficiently recognizable; we state this formally below.
\begin{observation}
It can be recognized whether a graph $G$ belongs to the family of graphs given in Theorem \ref{thm_cfn2} in $O(n^2)$ time.
\end{observation}

\section{Concluding remarks}
In this paper, we have furthered the study of connected forcing by characterizing graphs with connected forcing numbers 2 and $n-2$. In doing so, we employed novel combinatorial and graph theoretic techniques, which differ from the linear algebraic approaches typically used in deriving similar characterizations. A problem of interest is to obtain an analogous classification of graphs with connected forcing number or zero forcing number 3 and $n-3$; some of the techniques developed in the present paper could be useful toward that end. 

As part of our proof of Theorem \ref{thm_cfn2}, we introduced the notion of a connected forcing set which is required to contain a certain subset of the vertices of a graph (Definition \ref{restrained_forcing}). We will term this notion \emph{restrained connected forcing}; the notion of \emph{restrained zero forcing} can be defined analogously, i.e., a zero forcing set of $G=(V,E)$ restrained by $S\subset V$ is a zero forcing set which contains $S$. It would be interesting to study properties of the minimum zero forcing sets and the minimum connected forcing sets of a graph restrained by a given set $S$. Restrained forcing is at least as computationally hard as its unrestrained analogues, and could potentially lead to improved modeling of some of the physical phenomena related to the forcing process.

\section*{Acknowledgements}
This work is supported by the National Science Foundation, Grants No. 1450681, CMMI-1300477, and CMMI-1404864.

\end{document}